\documentclass[aps,showpacs,pra,amsmath,twocolumn]{revtex4}
\usepackage{graphicx}
\usepackage{bm}
\newcommand{\bra}[1]{\left\langle #1\right|}
\newcommand{\ket}[1]{\left| #1\right\rangle}
\newcommand{\avg}[1]{\langle #1\rangle}

\begin{document}                
\author{Wojciech Wasilewski}
\affiliation{Institute of Physics, Nicolaus Copernicus University, Grudziadzka 5, 87-100 Toru{\'n}, Poland}
\author{Konrad Banaszek}
\affiliation{Institute of Physics, Nicolaus Copernicus University, Grudziadzka 5, 87-100 Toru{\'n}, Poland}
\date{\today}
\title{Protecting an optical qubit against photon loss}%


\begin{abstract}
We consider quantum error-correction codes for multimode bosonic systems, such as optical fields, that are affected by amplitude damping. Such a process is a generalization
of an erasure channel. We demonstrate that the most
accessible method of transforming optical systems with the help of
passive linear networks has limited usefulness in preparing and manipulating such codes. These limitations stem directly from the recoverability condition for one-photon loss. We introduce a
three-photon code protecting against the first order of amplitude damping, i.e.\ a single photon loss,
and discuss its preparation
using linear optics with single-photon sources and conditional detection.
Quantum state and process tomography in the code subspace can be implemented using passive linear
optics and photon counting. An experimental proof-or-principle demonstration of elements of
the proposed quantum
error correction scheme for a one-photon erasure lies well within present technological capabilites.
\end{abstract}

\pacs{03.67.Pp, 42.50.Dv, 03.67.Hk}
\maketitle

\section{Introduction}
Quantum interference effects are susceptible to uncontrolled interactions with the environment, which can
undermine the advantage of quantum-enhanced information technologies. This difficulty was realized very early
in the development of quantum information processing, leading to the theory of quantum error correction (QEC)
\cite{KnillPRA97,QEC}. The basic idea underlying QEC is that if errors happen with a sufficiently small
probability or appear only in a restricted form, their effects can be suppressed by
preparing and processing
suitable robust states of multiplicated physical systems. Historically, attention was focused first
on ensembles of two-level systems --- physical qubits --- that interact independently with their environments
in an arbitrary way \cite{XYZ}. However, in certain implementations the dominant interaction with the
enviroment has a more specific form, enabling one to optimize the QEC strategy \cite{specialized}.

In this paper, we address amplitude damping leading to photon loss as the dominant decoherence mechanism in
photonic implementations of quantum information technologies. Particle loss, which is an analog of an erasure
in classical information theory, has been analyzed first by Grassl {\em et al.} \cite{GrasPRA97}, and by
Leung {\em et al.}\ \cite{ChuaLeunPRA97}, who took into account the bosonic statistics to derive codes
utilizing modal indistinguishability of particles. The presented examples of
codes used at least four photons, and they
have been exploited in a quantum memory proposal \cite{GingKokPRL03}. The use of mesoscopic superposition
states and the continuous-variable approach have also been studied \cite{Duze}.

A natural way to manipulate
photonic QEC codes is linear optics assisted with auxiliary photons and conditional detection. We demonstrate
here limitations of passive (i.e.\ not involving auxiliary conditional detection) linear optics networks for
this task. Further, we introduce a simple three photon code for encoding one logical qubit that is capable of
correcting for the loss of one photon. We derive a complete set of passive linear single-qubit gates, and
demonstrate the possibility to implement both state and process tomography with linear optics and photon
counting. This opens up prospects for proof-of-principle demonstrations of the three-photon code using
present experimental capabilities. We also present a conditional encoding circuit that maps an arbitrary
state of an input qubit in dual-rail representation onto the encoded subspace. Although its intrinsic success
rate is about $5\%$, we show that it can be boosted to the near-deterministic level using a suitably chosen
teleportation protocol. Finally, we present a conditional linear optical network that extends the available class
of single-qubit transformations, in particular enabling generation of an arbitrary superposition in the code subspace starting from
four photons.

\section{Photon loss codes}
Let us begin with a brief review of quantum error correction in the context of optical fields affected by amplitude damping.
Our goal is to shield a certain subspace in the Hilbert space of the system from errors described by a set
of noise operators $\{\hat F_i\}$. In the simplest case, the subspace is spanned by a pair of states which we will
denote by $\ket{L}$ and $\ket{H}$. An arbitrary state encoded as a superposition of $\ket{L}$ and $\ket{H}$ can be recovered after the action of the noise operators if and only if
the recoverability conditions are satisfied \cite{KnillPRA97,QEC}:
\begin{eqnarray}\label{Kn-La_conditions}
\avg{H|\hat F_i^\dagger \hat F_j|L}&=&0 \nonumber\\
\avg{H|\hat F_i^\dagger \hat F_j|H}&=&\avg{L|\hat F_i^\dagger \hat F_j|L}
\end{eqnarray}
for every pair of noise operators $\hat F_i$ and $\hat F_j$. Introducing a projector onto code subspace
$\hat{P}_{\cal C}=\ket{L}\bra{L}+\ket{H}\bra{H}$ the above conditions can be written in a compact
form as:
\begin{equation}
\hat{P}_{\cal C} \hat{F}^\dagger_i \hat{F}_j \hat{P}_{\cal C} = G_{ij} \hat{P}_{\cal C}
\end{equation}
with the help of coefficients $G_{ij}$. The above equation defines also higher-dimensional code subspaces, described by
projectors $\hat{P}_{\cal C}$ of rank higher than two.

Linear photon loss can be modeled as the transmission of a light beam through
a beam splitter with an amplitude transmission coefficient equal to $e^{-\gamma}$.
For a single light mode described by an annihilation operator $\hat{a}$, the transformation of
the input state $\hat{\varrho}$ is given by $\hat{\varrho} \rightarrow \sum_{n=0}^{\infty}
\hat{A}_n \hat{\varrho} \hat{A}_n^\dagger$, where $\hat{A}_n$ is a noise operator corresponding to
the loss of $n$ photons, given explicitly by \cite{ChuaLeunPRA97}:
\begin{equation}
\hat{A}_n=\frac{1}{\sqrt{n!}}(1-e^{-2\gamma})^{n/2}e^{-\gamma \hat a^\dagger \hat a}\,\hat a^n
\end{equation}
The leading order of an operator $\hat{A}_n$ in the loss parameter $\gamma$ is equal to ${\cal O}(\gamma^{n/2})$.
For a system of $N$ modes the noise operators are given by tensor products of the form
$\hat{A}_{n_1} \otimes \hat{A}_{n_2} \otimes \ldots \otimes \hat{A}_{n_N}$,where $n_j$ is the number
of photons removed from the $j$th mode.
When codes are constructed in a subspace characterized by a fixed number of $K$ photons, the noise operators corresponding to different total numbers $n_1 + n_2 + \ldots + n_N$ of lost photons
satisfy trivially the recoverability condition.

Let us now restrict our attention to the loss of at most one photon from the system. The noise operators
describing one-photon loss can be labeled with the index $i=1,\ldots,N$
of the mode in which the loss occurred.
If the losses affecting all the modes are identical with a single parameter $\gamma$, the one-photon loss operators can be written as:
\begin{equation}
\label{Eq:identicallosses}
\hat{F}_{i}
=
\sqrt{1-e^{-2\gamma}}e^{-\gamma (K-1)}\hat{a}_i,
\end{equation}
i.e.\ they are given by annihilation operators $\hat{a}_i$ of the respective modes
with identical multiplicative factors. Therefore a
necessary and sufficient condition for a recovery from one-photon loss is given by \cite{ChuaLeunPRA97}:
\begin{equation}
\label{Eq:Code} \hat{P}_{\cal C} \hat{a}^\dagger_i \hat{a}_j \hat{P}_{\cal C} = G_{ij} \hat{P}_{\cal C},
\;\;\; i,j = 1, \ldots, N
\end{equation}
where $(G_{ij})_{i,j=1,\ldots, N}$ are scalars forming a matrix $G$.

The tool for manipulating optical fields we focus on in this paper will be
linear transformations. Any such transformation of a
system of $N$ modes is described by a special unitary matrix $\Omega \in \textrm{SU}(N)$ with entries
$(\Omega_{ij})_{i,j=1,\ldots, N} $, which relates the input creation operators $\hat{a}_i^\dagger$ to the
output creation operators $\hat{b}_j^\dagger$ according to $ \hat{a}_i^\dagger = \sum_{j=1}^{N} \Omega_{ij}
\hat{b}_j^\dagger $. In a subspace of a fixed number of photons, spanned by multimode Fock states of the form
$|k_1 \ldots k_N \rangle = \prod_{i=1}^{N} (\hat{a}_i^\dagger)^{k_i} / \sqrt{k_i !} |\textrm{vac}\rangle$
with a constraint on the total number of excitations $\sum_{i=1}^{N} k_i = \textrm{const}$, the
transformation $\Omega$ induces a unitary representation $\hat{R}(\Omega)$ which can be obtained explicitly
by writing:
\begin{equation}
\hat{R}(\Omega)|k_1\ldots k_N\rangle = \prod_{i=1}^{N} \frac{1}{\sqrt{k_i !}} \left( \sum_{j=1}^{N}
\Omega_{ij} \hat{b}_j^\dagger \right)^{k_i}|\textrm{vac}\rangle
\end{equation}
and expanding the last expression in monomials of the form
$(\hat{b}_1^\dagger)^{l_1} \ldots
(\hat{b}_N^\dagger)^{l_N}/\sqrt{l_1!\ldots l_N!}$ acting on the vacuum state
$|\textrm{vac}\rangle$. The coefficients multiplying these expressions form columns of the representation
element $\hat{R}(\Omega)$.

\section{Limitations}
In this section we will show that the recoverability condition given in
Eq.~(\ref{Eq:Code}) restricts manipulations of photon loss codes that can be implemented with passive linear optics networks. First, let us demonstrate that no passive deterministic linear-optics network can encode a qubit carried by a
fixed number of photons, for example in the dual-rail representation. If the modes carrying the input qubit,
whose subspace is characterized by the projector $\hat{P}_{\text{in}}$, are combined with auxiliary modes,
prepared in a state described by a rank-one projector $\hat{P}_\text{aux}$, and subjected jointly
to a linear-optics transformation $\Omega$, the
resulting code subspace is given by $\hat{P}_{\cal C} = \hat{R}(\Omega) (\hat{P}_{\text{in}} \otimes
\hat{P}_\text{aux}) \hat{R}^\dagger(\Omega)$. However, the recoverability condition given in
Eq.~(\ref{Eq:Code}) applied to such a $\hat{P}_{\cal C}$ implies after a straightforward rearrangement that:
\begin{equation}
(\hat{P}_{\text{in}} \otimes \hat{P}_\text{aux}) \hat{a}_i^\dagger \hat{a}_j (\hat{P}_{\text{in}} \otimes
\hat{P}_\text{aux}) =(\Omega G \Omega^\dagger)_{ij} (\hat{P}_{\text{in}} \otimes \hat{P}_\text{aux})
\end{equation}
where we have used the fact that $\hat{R}^\dagger(\Omega) \hat{a}_{i}^\dagger \hat{R}(\Omega) = \sum_{j}
\Omega_{ji}^\ast \hat{a}^\dagger_j$. This means that the encoding in the subspace $\hat{P}_\text{in}$ is
itself robust against photon loss. An analogous argument shows that decoding cannot be deterministically
implemented with linear optics either.

The above reasoning shows that the robustness against photon loss is intrinsically connected to limitations
of possible linear-optics manipulations. Another important restriction can be shown when the matrix $G$ in
Eq.~(\ref{Eq:Code}) is proportional to identity, which is the case for a number of
examples presented in Ref.~\cite{ChuaLeunPRA97}. For such codes no
continuous set of unitary transformations on the encoded qubit can be realized deterministically using
passive linear optics. In order to prove this statement, let us assume the contrary. The set of unitary transformations
inducing gates on the encoded qubit forms then a continuous subgroup of $\textrm{SU}(N)$. This implies the
existence of an element in the Lie algebra $\Lambda \in \mathrm{su}(N)$ such that for any real $\alpha$ the
unitary transformation ${\Omega}(\alpha) = \exp(i\alpha\Lambda)$ preserves the code subspace defined by the
projector $\hat{P}_{\cal C}$. This can be written as $ \hat{R}(\Omega(\alpha))\hat{P}_{\cal C} =
\hat{P}_{\cal C} \hat{R}(\Omega(\alpha))\hat{P}_{\cal C} $. Differentiating this expression with respect to
$\alpha$ and inserting $\alpha = 0$ yields:
\begin{equation}\label{Eq:RP=PRP}
\hat{R}(\Lambda)\hat{P}_{\cal C} = \hat{P}_{\cal C} \hat{R}(\Lambda)\hat{P}_{\cal C}
\end{equation}
However, $\hat{R}(\Lambda)$ is a linear combination of the operators $\hat{a}^\dagger_i \hat{a}_j +
\hat{a}^\dagger_j \hat{a}_i$, $i(\hat{a}^\dagger_i \hat{a}_j - \hat{a}^\dagger_j \hat{a}_i)$ and
$\hat{a}^\dagger_i \hat{a}_i - \hat{a}^\dagger_j \hat{a}_j$ with $i,j=1,\ldots, N$ and $i\neq j$, which span
the representation of the algebra $\mathrm{su}(N)$ in the space of $N$ bosonic modes \cite{Gilmore}.
For codes satisfying $G_{ij} \propto \delta_{ij}$ the error-correcting
condition given in Eq.~(\ref{Eq:Code}) immediately implies
that $\hat{P}_{\cal C} \hat{R}(\Lambda)\hat{P}_{\cal C} = 0$. Consequently the left-hand side of
Eq.~(\ref{Eq:RP=PRP}) vanishes as well. This means that the action of the unitary transformations
${\Omega}(\alpha)$ in the code subspace $\hat{P}_{\cal C}$ is trivial and
$\hat{R}(\Omega(\alpha))\hat{P}_{\cal C} = \hat{P}_{\cal C}$. The reasoning presented above demonstrates that
linear optics transformations on the $N$ modes of the code can generate at most a discrete group of
single-qubit gates in the encoded subspace.

\section{Three-photon code}

We will now present a three-photon code that protects one logical qubit
against a photon loss. The two logical states $|L\rangle$ and
$|H\rangle$ of the encoded qubit are given by the following
three-photon states:
\begin{eqnarray}
|L\rangle & = & \frac{1}{\sqrt{3}}(|003\rangle + |030\rangle + | 300 \rangle) \nonumber \\
|H\rangle & = & | 111 \rangle. \label{Eq:3photoncode}
\end{eqnarray}
It is straightforward to verify that the recoverability condition from Eq.~(\ref{Eq:Code}) is satisfied for
the three photon code with the matrix $G$ equal to identity. Therefore both of
the general observations proven in the preceding section apply to our code. Assuming the availability of single photon sources,
a logical state that can be generated easily is $H=|111\rangle$. The first step towards an experimental realization
of the code would be the generation of other states from the code subspace. One way to achieve this is to construct
single gates operating in the code subspace, and we will discuss this approach within the linear-optics paradigm.

Let us first consider a scenario when the three signal modes carrying the encoded qubit are
combined with an arbitrary number of auxiliary vacuum modes and subjected jointly to a unitary linear
transformation. The output in the three signal modes is accepted conditionally on the detection of zero
photons in all the auxiliary modes. This procedure includes deterministic gates as a specific case.
If the auxiliary modes are prepared and detected in the vacuum state, the transformation of the state
contained in the signal modes depends only on a $3\times 3$ sector of the entire unitary matrix describing
the transformation of the modes, which relates the input signal modes $\hat{a}_1^\dagger, \hat{a}_2^\dagger,
\hat{a}_3^\dagger$ to the output signal modes $\hat{b}_1^\dagger, \hat{b}_2^\dagger, \hat{b}_3^\dagger$. We
will denote this sector by $\tilde{\Omega} = (\Omega_{ij})_{i,j=1,2,3}$, and the resulting conditional
transformation acting in the space of the three signal modes by $\hat{R}(\tilde{\Omega})$. The sector
$\tilde{{\Omega}}$ can be extended to a unitary matrix by adding auxiliary modes provided that the inequality
$\tilde{{\Omega}}^\dagger \tilde{{\Omega}} \le \hat{\openone}$ is satisfied, and this condition can be always
met by an appropriate rescaling of $\tilde{{\Omega}}$ at the cost of lowering the success rate of the gate.

We will now derive constraints on $\tilde{\Omega}$ that result from the condition
$\hat{R}(\tilde{\Omega})\hat{P}_{\cal C} = \hat{P}_{\cal C} \hat{R}(\tilde{\Omega})\hat{P}_{\cal C}$ stating
that the code subspace needs to be preserved. We will discuss separately three cases according to the number
of zeros in the first column of $\tilde{\Omega}$. Let us first assume that the product
$\omega_{11}\omega_{21}\omega_{31}\neq 0$. In this case we can write:
\begin{equation}
\tilde{\Omega} = \left(
\begin{array}{ccc}
\omega_{11} & 0 & 0 \\
0 & \omega_{21} & 0 \\
0 & 0 & \omega_{31}
\end{array}
\right) \left(
\begin{array}{ccc}
1 & x & u \\
1 & y & v \\
1 & z & w
\end{array}
\right)
\end{equation}
The condition $\langle 300 | \hat{R}(\tilde{\Omega}) | H \rangle = \langle 030 | \hat{R}(\tilde{\Omega}) | H
\rangle = \langle 003 | \hat{R}(\tilde{\Omega}) | H \rangle$ implies then that $z=(xy)^{-1}$ and
$w=(uv)^{-1}$. Furthermore, the requirement $\langle 210 | \hat{R}(\tilde{\Omega}) | H \rangle = \langle 120
| \hat{R}(\tilde{\Omega}) | H \rangle = 0$ gives $y=\exp(\pm2\pi i/3)x$ and also that $x^3 = 1$. This means
that the parameters $x,y,z$ are arbitrarily permuted cube roots of one $\{ 1, \exp(\pm2\pi i/3)\}$. An
analogous consideration applied to the scalar products of $ \hat{R}(\tilde{\Omega}) | H \rangle$ with
$\langle 2 0 1|$ and $\langle 1 0 2|$ implies that also $u,v,w$ must be equal to three different cube roots
of one. Finally, the condition $\langle 012 | \hat{R}(\tilde{\Omega}) | H \rangle = 0$ gives $u^\ast x +
v^\ast y + w^\ast z = 0$, which means that the vectors $(x,y,z)$ and $(u,v,w)$ are orthogonal.

If we demand that $\hat{R}(\tilde{\Omega})|L\rangle$ also remains in the code subspace, then vanishing projections
onto states $\langle 120|$ and $\langle 102|$ imply that $x^\ast \omega_{11}^3 + y^\ast \omega_{21}^3 +
z^\ast \omega_{31}^3 = u^\ast \omega_{11}^3 + v^\ast \omega_{21}^3 + w^\ast \omega_{31}^3 = 0$. This means
that the vector $(\omega_{11}^3, \omega_{21}^3, \omega_{31}^3)$ is orthogonal to both $(x,y,z)$ and
$(u,v,w)$, and the only nontrivial possibility to meet this condition is $\omega_{11}^3 = \omega_{21}^3 =
\omega_{31}^3$. Thus, the parameters $\omega_{11}$, $\omega_{21}$, and $\omega_{31}$ are given, up to a
common multiplicative constant, by arbitrary cube roots of one. The upper bound on their magnitude can be
derived from the condition $\tilde{\Omega}^\dagger \tilde{\Omega} \le \hat{\openone}$, giving $|\omega_{11}|
= |\omega_{21}| = |\omega_{31}| \le 1/\sqrt{3}$. The success rate of the gate performed on the encoded qubit
is maximized when the last inequality is saturated. Then the matrix $\tilde{\Omega}$ is unitary, and there is
no need to introduce auxiliary vacuum modes.

The second case is when exactly one element in the first column of $\tilde{\Omega}$ is zero. Because of the
symmetry of the code, we can assume with no loss of generality that $\Omega_{11}= 0$. Considering the scalar
products $\langle 2 1 0 | \hat{R}(\tilde{\Omega}) |H\rangle$ and $\langle 2 0 1 | \hat{R}(\tilde{\Omega})
|H\rangle$ shows that also $\Omega_{12} = \Omega_{13} = 0$. This means that $\hat{R}(\tilde{\Omega})|H\rangle
= 0$, and it is sufficient to ensure that $\hat{R}(\tilde{\Omega})|L\rangle$ stays in the code subspace.
Because of conditions $\langle 2 1 0 |\hat{R}(\tilde{\Omega})|L\rangle = 0$ and $\langle 1 2 0
|\hat{R}(\tilde{\Omega})|L\rangle = 0$ both the elements of the pairs $(\Omega_{22}, \Omega_{32})$ and
$(\Omega_{22}, \Omega_{32})$ must be either zero or non-zero. In the former case it is straightforward to see
that $\langle H |\hat{R}(\tilde{\Omega})|L\rangle = \langle L |\hat{R}(\tilde{\Omega})|L\rangle  = 0$. In the
latter case, considering vanishing projections of $\hat{R}(\tilde{\Omega})|L\rangle$ onto $\langle 2 1 0|$,
$\langle 1 2 0|$, $\langle 201|$, $\langle 1 0 2$, $\langle 012|$, and $\langle 021|$ gives that
$\Omega_{21}/\Omega_{31} = \Omega_{22}/\Omega_{32}= \Omega_{23}/\Omega_{33} = 1$, which again implies that
$\langle H |\hat{R}(\tilde{\Omega})|L\rangle = \langle L |\hat{R}(\tilde{\Omega})|L\rangle  = 0$.

Thus we are left with the case when at least two elements in the first column of $\tilde{\Omega}$ are zero.
Applying the reasoning from the preceding paragraph to the second and the third column shows that each
column must have at least two zero elements. Considering the remaining cases leads to a conclusion that
$\tilde{\Omega}$ must be given, up to a permutation of rows and columns, to a diagonal matrix with elements
proportional to arbitrarily chosen cube roots of one. The success rate is largest when the proportionality
factor has unit absolute value, giving a unitary form of $\tilde{\Omega}$. As the code words are symmetric in
the modes, the matrices $\tilde{\Omega}$ derived above can be considered up to arbitrary permutations of rows
and columns, which obviously do not alter transformations in the code subspace.

\begin{figure}
\center
  \includegraphics[scale=0.3]{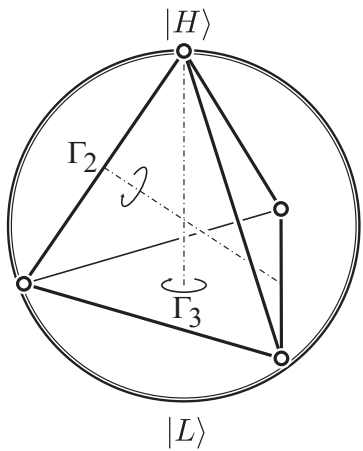}
  \caption{The Bloch sphere of the encoded qubit. The logic state $\ket{H}$ can be transformed
deterministically using linear optics into any of three states indicated with dots. All linear optics
transformations that do not involve auxiliary photons are given by the rotation group of the tetrahedron,
generated from a $2\pi/3$ rotation $\Gamma_3$ around the vertical axis and a $\pi$ rotation $\Gamma_2$ around
the axis passing through the midpoints of the opposite edges of the tetrahedron.}\label{fig:Sphere}
\end{figure}

Consequently, the entire set of $\tilde{\Omega}$s that preserve the code subspace consists of unitary
matrices. This set can be generated from two transformations, given up to overall phase factors by:
\begin{equation}
\Gamma_2 =\frac{1}{\sqrt{3}}\begin{pmatrix}
1 & 1 & 1 \\
1 & \zeta & \zeta^2 \\
1 & \zeta^2 & \zeta \end{pmatrix}, \qquad \Gamma_3 = \begin{pmatrix}
1 & 0 & 0 \\
0 & 1 & 0 \\
0 & 0 & \zeta \end{pmatrix}
\end{equation}
where $\zeta=\exp(2\pi i /3)$. The above linear-optics
transformations are respectively a tritter \cite{ZukoZeilPRA97}
splitting equally any input port between the three output ports, and
a simple $2\pi/3$ phase shift applied to one of the modes. The geometric
structure of the set of allowed transformations can be understood by
considering logic states that are obtained from the basis state
$|H\rangle$ by a repetitive application of the generators $\Gamma_2$
and $\Gamma_3$ in an arbitrary order. It is easy to verify that this
yields three states of the form $|T_l\rangle = (|H\rangle -
\sqrt{2}\zeta^{l}|L \rangle)/\sqrt{3}$ where $l =1,2,3$, which
together with $|H\rangle$ form a regular tetrahedron in the Bloch
sphere of the logic qubit, shown in Fig.~\ref{fig:Sphere}. Thus the
entire set of single-qubit operations that can be implemented using
linear optics without auxiliary photons is identical with the
rotation group ${\cal T}$ of that tetrahedron, and the generators
$\Gamma_2$ and $\Gamma_3$ correspond to two rotations by $\pi$ and
$2\pi/3$ about axes depicted in Fig.~\ref{fig:Sphere}.

The measurement of the logical qubit in the basis $\{|L\rangle,
|H\rangle\}$ can be implemented by counting photon numbers in the
three modes, which destructively yields both the qubit value in the computational
basis and the error.
Preceding photon counting by one of the deterministic
linear-optics gates  $\Gamma_3\Gamma_2$, $\Gamma_3^2\Gamma_2$, or $\Gamma_2$,
allows one to realize a projection in the basis
composed of any of the states $|T_l\rangle$ and its orthogonal
complement. This provides sufficient data for a complete tomographic
reconstruction \cite{OptimalTomo}. Furthermore, given a quantum process that
preserves the code subspace, its action can be fully characterized by feeding it
with states $\ket{H}, \ket{T_l}$, and performing state tomography on the output.

\begin{figure} \center
  \includegraphics[scale=0.75]{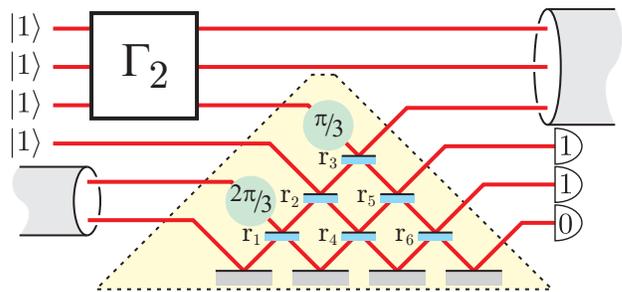}
  \caption{
A universal circuit for encoding a single qubit in the dual-rail
representation using four auxiliary photons. Three of the auxiliary photons
are sent through a tritter $\Gamma_2$ to generate the state $\ket{T_3}$.
One of the output modes from the tritter is combined with the remaining photon and the input qubit.
The output is accepted upon detecting a combination
$110$. Filled circles denote phase shifts, and horizontal bars are beam splitters with
real amplitude transmission and reflection coefficients. The convention is that the
beam reflected off the thickened side of a beam splitter acquires a minus sign.
The sign change is introduced also by four mirrors. The amplitude reflection
coefficients for the beam splitters are: $r_1 = 0.586$, $r_2 = 0.728$, $r_3 = 0.448$,
$r_4 = 0.625$, $r_5 = 0.837$, and $r_6 = 0.984$.}\label{fig:encoding}
\end{figure}

\section{Universal encoding}

In this section we will present a universal linear-optics circuit that converts a single qubit in the standard
dual-rail representation into a qubit encoded in the loss-proof $\{|L\rangle, |H\rangle \}$ basis. As we
demonstrated in this paper, such a procedure cannot be implemented derministically using a passive
linear-optics network. Our circuit, depicted in Fig.~\ref{fig:encoding}, combines the input qubit carried by
one photon in a superposition of two modes with four auxiliary single photons, and conditions the output upon
the detection of the sequence $|110\rangle$ in the three monitored modes. The three remaining output modes
that leave the circuit contain then the encoded qubit.

The circuit begins with the preparation of the state
$\ket{T_3}=(|H\rangle - \sqrt{2}|L \rangle)/\sqrt{3}$ from three of
the auxiliary photons using a tritter $\Gamma_2$. One of the modes
leaving the tritter, which we shall label with the index $s$, is
then combined with the input qubit and the remaining auxiliary
photon in a four-mode network encircled with a dashed line in
Fig.~\ref{fig:encoding}. The purpose of this network is to modulate
the probability amplitudes of the Fock states $\ket{0}_s$,
$\ket{1}_s$ and $\ket{3}_s$ of the mode $s$ depending on the state
of the input qubit. The action of the network can be written as:
\begin{align}
\ket{n}_s \ket{1} \ket{10} & \rightarrow c_{ln} \ket{n}_s \ket{110} +\ldots\nonumber \\
 \ket{n}_s \ket{1} \ket{01} & \rightarrow c_{hn} \ket{n}_s \ket{110} +\ldots
\end{align}
where the dots denote terms with combinations of Fock states in the
detected modes other than $\ket{110}$. In order to map an arbitrary
state $\alpha \ket{10} + \beta \ket{01}$ of the input state onto the
encoded state $\alpha \ket{L} + \beta \ket{H}$, the coefficients
$c_{ln}$ and $c_{hn}$ need to satisfy, up to an overall phase
factor, the following equations:
\begin{align}
c_{l0} = c_{l3} & =  \sqrt{3p/2} \nonumber \\
c_{h1} & = - \sqrt{3p} \label{Eq:clnchn} \\
c_{h0} = c_{l1} = c_{h3} & = 0 \nonumber
\end{align}
where $p$ is the overall success rate of the encoding circuit. It is
straightforward to verify by numerical means that these constraints are indeed
fulfilled for a network described in Fig.~\ref{fig:encoding}.

The overall success rate of the encoding circuit presented above is
equal to $p=4.86\%$. We performed a numerical search over networks
that combine in an arbitrary way all the six input modes and
any number of additional vacuum modes, which did not improve the
success rate. The presented encoding circuit can be however applied
to encode a single copy of a qubit in an unknown state by using it
``off-line'', following the idea of teleportation-based
computational primitives \cite{Primitives,KLM}. The first
step is to apply the encoding circuit to a qubit
that belongs to a maximally entangled pair composed of two photons
in a state $|\psi_+\rangle = (\ket{0101} +
\ket{1010})/\sqrt{2}$, repeating the procedure on freshly prepared pairs
until the successful operation is achieved.
Then the second qubit from the successful pair is measured jointly
with the input qubit using a projection onto four maximally entangled states
defined as:
\begin{equation}
(\hat{U}_g \otimes \hat{\openone})\ket{\psi_+} ,
\; \; \; g \in {\cal T}',
\end{equation}
where ${\cal T}' \subset {\cal T}$ is a subgroup of tetrahedron rotations consisting of
the identity and three rotations by $\pi$, and $\hat{U}_g$ are unitaries
that generate these transformations in the single-qubit Bloch sphere. The
measurement step can be performed nearly deterministically using
universal linear-optics quantum circuitry with auxiliary single
photons and fast feed-forward \cite{KLM}. This leaves the
encoded qubit in the desired state up to one of the rotations from the subgroup
${\cal T}'$ which can be compensated for by an optional application
of one of the deterministic
linear-optics transformations $\Gamma_2$, $\Gamma_3\Gamma_2\Gamma_3^2$, or
$\Gamma_3^2\Gamma_2\Gamma_3$.

\section{Conditional single qubit operations}

\begin{figure}
\center
  \includegraphics[scale=0.9]{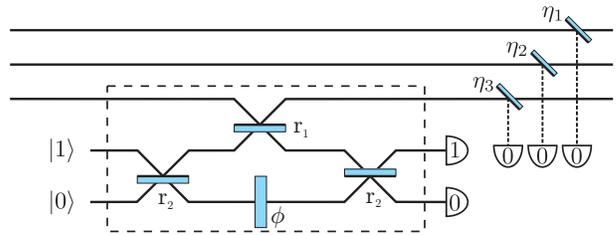}
  \caption{A network for conditional manipulations of the three-photon code. The three upper modes carry the
encoded qubit. One of the modes undergoes a linear transformation with two auxiliary modes carrying
respectively one and zero photons, shown within a dashed box. The transformation network involves beam
splitters with amplitude reflectivities $r_1$ and $r_2$ and a phase delay $\phi$. The output is accepted when
the auxiliary modes generate one- and zero photon events on the detectors. The three signal modes are then
attenuated by beam splitters with complex amplitude transmissivities $\eta_1$, $\eta_2$ and $\eta_3$, and the
gate is successful when none of the detectors monitoring the reflected beams registers any
photons.}\label{fig:CPG}
\end{figure}

In the remaining part of this paper we will present a class of single-qubit transformations that can be
realized with one auxliliary photon, depicted schematically in Fig.~\ref{fig:CPG}. Let us first
consider the network within a
dashed box of Fig.~\ref{fig:CPG}. This circuit combines one of the signal modes with two auxiliary modes prepared
respectively in a one- and a zero-photon state. The output is accepted when the detectors monitoring
auxiliary modes detect respectively one and zero photons. The principle of operation resembles that of the
nonlinear phase shift gate \cite{PhaseShift}. An easy calculation shows that a superposition $\alpha_0
\ket{0} + \alpha_1 \ket{1} + \alpha_3 \ket{3}$ of the signal mode is conditionally transformed into $
\alpha_0 c_0 \ket{0} + \alpha_1 c_1 \ket{1} + \alpha_3 c_3 \ket{3} $ where the coefficients $c_0$, $c_1$, and
$c_3$ are given by:
\begin{eqnarray}
c_0&=&-r_1r_2^2 +e^{i\phi}(1-r_2^2)\nonumber\\
c_1&=&-c_0r_1 +r_2^2(r_1^2-1)\nonumber\\
c_3&=&r_1^2[-c_0r_1 +3r_2^2(r_1^2-1)]
\end{eqnarray}
and $r_1, r_2, r_3$ and $\phi$ are respectively the reflection coefficients and the phase shift as depicted
in Fig.~\ref{fig:CPG}. The above coefficients have a simple physical interpretation. Two paths contribute to
$c_0$: either auxiliary photon bounces of all the beamsplitters, or it goes through both lower ones. For
$c_1$ either the same happens while the signal photon bounces off the upper beamsplitter, or the photons
cross their ways. In case of $c_3$ either all signal photons bounce off the upper beamspliter, or one of
them exchanges with the auxiliary one.

When the above network is applied to one of the signal modes carrying the three-photon code as shown in
Fig.~\ref{fig:CPG}, an arbitrary state of the encoded qubit $|\psi\rangle = \alpha_L \ket{L} + \alpha_H
\ket{H}$ is mapped onto:
\begin{equation}
\ket{\psi} \rightarrow \frac{\alpha_L}{\sqrt{3}} ( c_3 \ket{003} + c_0 \ket{030} + c_0 \ket{300} ) + \alpha_H
c_1 \ket{111}.
\end{equation}
Different weights introduced by the coefficients $c_0$ and $c_3$ in the basis state $\ket{L}$ can be
equalized by inserting beam splitters characterized by complex amplitude transmission coefficients $\eta_1$,
$\eta_2$, and $\eta_3$, and accepting the output if no photons are rerouted to detectors monitoring
reflections from the beam splitters. The transmission coefficients $\eta_1$, $\eta_2$, and $\eta_3$ are
chosen depending on the ratio $\kappa = c_3/c_0$. When $|\kappa| \ge 1$, the states contributing to the logic
state $\ket{L}$ can be balanced by attenuating the third mode by $\eta_3 = \kappa^{-1/3}$ and leaving the
remaining modes intact with $\eta_1 = \eta_2 = 1$. In the opposite case, when $|\kappa| < 1$, the first two
modes need to be attenuated with $\eta_1 = \eta_2 = \kappa^{1/3}$, with the third mode fully transmitted with
$\eta_3 = 1$.

\begin{figure}
\center  \includegraphics[scale=0.8]{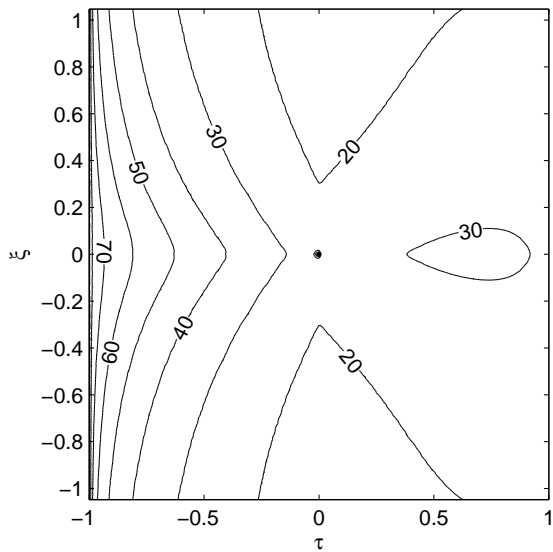}
  \caption{A contour plot of the maximal attainable success rate $S= ||\hat{A}||^2$, expressed in percent, of the conditional transformation of the logic qubit given by an operator $\hat{A}$ parameterized according to Eq.~(\ref{Eq:Aparam}). }\label{fig:BSe}
\end{figure}

The resulting transformation of the logic qubit $\ket{\psi}$ can be represented as a transformation
$\ket{\psi} \rightarrow \hat{A} \ket{\psi}$, where the operator $\hat{A}$ is given in the logic basis $\{
\ket{L}, \ket{H} \}$ as a diagonal matrix $\textrm{diag}(c_0, c_1 \kappa^{-1/3})$ if $|\kappa|\ge 1$ and
$\textrm{diag}(c_3, c_1 \kappa^{2/3})$ if $|\kappa|<1$. We will characterize the success rate $S$ of the
conditional operation $\ket{\psi} \rightarrow \hat{A} \ket{\psi}$ by the squared operator norm $S =
||\hat{A}||^2$. The renormalized operator $\hat{A}/||\hat{A}||$ can be parameterized as
\begin{equation}
\label{Eq:Aparam} \hat{A}/||\hat{A}|| = \textrm{diag} (\sqrt{1-\tau}, e^{i\xi}
\sqrt{1+\tau})/\sqrt{1+|\tau|},
\end{equation}
 where the relative phase $\xi$ can be restricted to
the range  $-\pi/3 \le \xi \le \pi/3$, as phase gates that are multiples of $2\pi/3$ can be implemented
deterministically, and the physical range of $\tau$ is $-1 \le \tau \le 1$, with $\tau = -1$ and $\tau=1$
corresponding to unnormalized projections onto $\ket{L}$ and $\ket{H}$. In Fig.~\ref{fig:BSe} we present the
maximum success rate $S$ of a gate as a function of its parameters $\tau$ and $\xi$, obtained from numerical
optimization of the success rate over the physical parameters $r_1$, $r_2$, and $\phi$ of the network
depicted in Fig.~\ref{fig:CPG} under the constraint of fixed $\tau$ and $\phi$. For $\tau=-1$, when the gate
projects onto $|L\rangle$, one obtains a simple and intuitive solution $r_1=1/\sqrt{2}$ and $r_2=1$, implying
that the gate relies on the Hong-Ou-Mandel two-photon interference effect. If the signal mode contains one
photon, the conditioning detector will register either zero or two photons, thus blocking the $\ket{H}$
component in the input state.

One can consider a more general gate which combines all the signal modes with one single-photon mode and an
arbitrary number of vacuum modes, and implements an general linear-optics transformation with the output
accepted upon the detection of the auxiliary modes containing exactly one photon. We have used techniques
developed in this paper to optimize numerically the success rate of such a gate on a $5\times 6$ grid in the
($\xi$,$\tau$) plane. As the optimization problem is highly nonlinear, we restarted the maximization
procedure several hundred times at each grid point with random initial conditions. However, the numerical
search yielded no gate that would outperform the one presented above, which strongly hints towards its
optimality. Further insights could be perhaps gained by exploring approach developed by Eisert \cite{Jens}.

\section{Conclusions}

In conclusion, we demonstrated general limitations of passive linear-optics manipulations of quantum
error correcting codes for photon loss. These restrictions are intimately linked to the recoverability
condition itself. Further, we proposed a three photon code whose preparation and properties could be
tested using current experimental capabilities \cite{4fotonyexp}. A remaining open problem is how to implement
within the linear-optics paradigm the non-destructive
determination of the error syndrome and a subsequent correction of the encoded qubit.
Nevertheless, the set of tools developed in this paper could be used for a proof-of-principle
demonstration that quantum coherence remains preserved in the code subspace despite photon loss.

In some applications, error-correcting codes for amplitude damping might provide an alternative to quantum repeaters
\cite{QRepeaters}, which require two-way classical communication and intermediate stations equipped with quantum memories. However, it would be difficult to give a general comparison between these two approaches, as they use distinct types of resources. Finally, let us note that the codes for amplitude damping could compensate for losses in the transmission channel as well as at the detection stage, when received states are measured with photodetectors preceded by passive linear network. This could be perhaps used to compensate for detector inefficiencies in some protocols involving quantum correlations.

We acknowledge insightful discussions with D. Gottesman, M. Horodecki, R. Laflamme, D. Leung,
N. L\"{u}tkenhaus, I. A. Walmsley, and
M. \.{Z}ukowski. This work has been supported by the European Commission under the Integrated Project Qubit
Applications (QAP) funded by the IST directorate as Contract Number 015848, Polish MNiSZ grant
1~P03B~011~29 and AFOSR under grant number FA8655-06-1-3062.

\end{document}